\documentclass[11pt]{article}
\usepackage[utf8]{inputenc}
\usepackage[T1]{fontenc}
\usepackage{lmodern}
\usepackage[margin=1in]{geometry}
\usepackage{amsmath,amssymb}
\usepackage{graphicx}
\usepackage{booktabs}
\usepackage{hyperref}
\usepackage{xcolor}
\usepackage{natbib}
\usepackage{enumitem}
\usepackage{placeins}
\usepackage{url}
\usepackage{microtype}
\hypersetup{colorlinks=true, linkcolor=blue!70!black, citecolor=blue!70!black, urlcolor=blue!70!black}
\newcommand{\Tdec}{T_{\mathrm{dec}}}
\newcommand{\bstar}{b^\star}

\title{When Is a Columnar Scan Bandwidth-Bound?\\ A Decode-Throughput Law and Its Cross-Hardware Validation}
\author{Madhulatha Mandarapu\thanks{madhulatha@samyama.ai} \and Sandeep Kunkunuru\thanks{sandeep@samyama.ai}}
\date{VaidhyaMegha Private Limited, India\\[2pt]\url{https://samyama.ai/}\\[8pt]June 2026}

\begin{document}
\maketitle

\begin{abstract}
A columnar scan that decompresses, filters, and aggregates \emph{should} be limited only by memory
bandwidth---the roofline floor $T \ge \text{BytesRead}/\beta$---yet real kernels are often
compute-bound and leave bandwidth idle. We give a predictive answer to \emph{when} a scan is
bandwidth-bound. Across encodings, predicate selectivities, and two different machines, we find
that a decoder's \emph{value throughput} $\Tdec$ (values decoded per second) is essentially
independent of the bit-width $b$: it is set by the decode \emph{layout/strategy}, not by how many
bits each value occupies. Consequently the achieved bandwidth fraction obeys a one-parameter law,
$f = \min\!\big(1,\ \Tdec\, b / (8\beta)\big)$, with the compute-to-bandwidth ridge at
$\bstar = 8\beta/\Tdec$. Fitting one $\Tdec$ per strategy reproduces measured bandwidth fractions
with median error $0.027$ on x86/AVX2 and $0.003$ on a held-out Apple~M4/NEON machine, and the ridge
$\bstar$ shifts correctly with each machine's bandwidth. Inserting FastLanes' reported decode
throughput into the law reproduces its ``decode is free at three bits'' headline as the
large-$\Tdec$ limit, unifying our decoder and hand-tuned production code in one curve. We
add two crossovers, validated on both machines: branch-free predicate evaluation beats branchy in a
mid-selectivity band (the $\sigma(1-\sigma)$ misprediction parabola), and zone-map skipping
is \emph{clustering}-gated rather than selectivity-gated. We release the micro-benchmark, the
correctness oracle, and a one-command reproduction. This is a baseline and model, not a faster
kernel: our portable C decoders reach $\sim 2$ values/cycle, far below hand-tuned SOTA; the law
holds because it is parameterized by the measured $\Tdec$.
\end{abstract}

\section{Introduction}
Analytical query engines spend much of their time scanning compressed columns. The folklore goal is
to make such scans \emph{bandwidth-bound}: limited by how fast bytes stream from DRAM, not by decode
or predicate instructions. The roofline model~\citep{williams2009roofline} makes this precise---a
kernel of operational intensity $I$ (useful work per byte) attains $\min(\pi,\beta I)$, and is
bandwidth-bound iff $I$ is below the ridge $\pi/\beta$. Modern columnar systems
\citep{boncz2005monetdb,lang2016datablocks,kuschewski2023btrblocks,afroozeh2023fastlanes} push decode
below this ridge with SIMD-friendly layouts. But the field lacks a simple, \emph{predictive} answer
to the operational question a practitioner actually asks: \emph{given my encoding, my selectivity,
and my hardware, what fraction of memory bandwidth will this scan achieve, and at what bit-width does
decode stop being free?}

The phenomena are individually known. Selectivity-dependent scan throughput and zone-map skipping are
measured by Data Blocks~\citep{lang2016datablocks} and, recently and thoroughly, by
\citet{zeng2023empirical}; branch-free vs branchy SIMD scans are studied by
\citet{polychroniou2015lightweight}. What is missing is a closed-form model that ties them to the
roofline and \emph{transfers across hardware}. This paper supplies one.

\paragraph{Contributions.}
\begin{itemize}[leftmargin=1.4em,topsep=2pt,itemsep=1pt]
  \item \textbf{A bandwidth-fraction law (\S\ref{sec:law}).} The empirical observation that decode
  value-throughput $\Tdec$ is bit-width-independent yields $f=\min(1,\Tdec b/(8\beta))$ with ridge
  $\bstar=8\beta/\Tdec$. One parameter per decode strategy.
  \item \textbf{Cross-hardware validation (\S\ref{sec:results}).} Median fit error $0.027$ on
  x86/AVX2 and $0.003$ on held-out Apple~M4/NEON; the ridge moves correctly with $\beta$. The law
  reproduces FastLanes' ``decode is free'' result as its $\Tdec\!\to\!\infty$ limit.
  \item \textbf{Two crossovers (\S\ref{sec:cross}).} The branch-free/branchy mid-selectivity band
  (predicted within $\pm1$ selectivity decade) and the result that zone-map skipping is
  \emph{clustering}-gated, sharpening the ``low-selectivity only'' folklore.
  \item \textbf{An open, reproducible harness} with a correctness oracle and negative controls; one
  command regenerates every number.
\end{itemize}

\paragraph{What this is not.} We do not beat FastLanes or BtrBlocks. Our portable-C decoders reach
$\sim\!2$ values/cycle, well below FastLanes' hand-tuned $\sim\!40$; we \emph{model the regime} they
operate in. The law is parameterized by the measured $\Tdec$, so the science does not depend on
matching SOTA absolute speed (\S\ref{sec:limitations}).

\section{Problem and Model}\label{sec:model}
A scan reads a compressed column, decodes each value, evaluates a predicate, and aggregates. Let
$\beta$ be sustainable memory bandwidth (bytes/s) and $b$ the bits per (encoded) value. Reading a
column of $N$ values costs $\text{BytesRead}=Nb/8$, so the roofline floor is $T \ge Nb/(8\beta)$ and
the achieved \emph{bandwidth fraction} is
\begin{equation}
  f \;=\; \frac{\text{BytesRead}/T}{\beta} \;=\; \frac{\text{achieved read GB/s}}{\beta} \in (0,1].
\end{equation}
Write $\Tdec$ for the decoder's value throughput (values/s) in the compute-bound regime, i.e. the
rate at which the kernel can produce decoded values when bandwidth is not the bottleneck. Then the
achieved read bandwidth is $\Tdec \cdot b/8$ bytes/s until it saturates $\beta$:
\begin{equation}\label{eq:law}
  \boxed{\,f \;=\; \min\!\Big(1,\ \frac{\Tdec\, b}{8\beta}\Big)\,}, \qquad
  \bstar \;=\; \frac{8\beta}{\Tdec}\ \ \text{(the compute$\to$bandwidth ridge).}
\end{equation}
Equation~\eqref{eq:law} is the roofline written in scan-native units. Its predictive content is the
empirical claim, validated in \S\ref{sec:results}, that \textbf{$\Tdec$ is independent of $b$}: a
decoder processes a near-constant number of values per second regardless of how many bits each value
takes, so the \emph{bytes} it pulls---and thus $f$---grow linearly in $b$ until the ridge $\bstar$.

\section{Method: encodings, kernels, harness}\label{sec:law}
We implement three decode strategies over 32-bit integer columns:
(i)~\textbf{scalar row-major} bit-unpacking (a serial variable-shift extractor; the compute-bound
baseline); (ii)~a \textbf{FastLanes-style transposed} layout~\citep{afroozeh2023fastlanes} that stores
1024-value blocks across 32 lanes so that all lanes share one shift+mask and the inner loop
auto-vectorizes; and (iii)~\textbf{byte-aligned} widths ($b\in\{8,16,32\}$) that coincide with plain
integer arrays. Predicates are equality counts and a materializing select; we also implement RLE and
per-block zone maps (min/max skipping). The fused kernels decode, filter, and aggregate in one pass.

$\beta$ is measured per machine as STREAM-Triad~\citep{mccalpin1995stream} and an empirical
sequential-read maximum (buffers $\gg$ LLC); we report $f$ against the empirical read maximum (tighter
than theoretical peak). A correctness oracle checks encode/decode round-trips for all widths $1\!-\!32$
in both layouts, scan counts against brute force, and three negative controls (byte-aligned $=$
generic decode; random data $\Rightarrow$ no zone skipping; sub-LLC throughput $>\beta$). All code,
seeds, and a one-command reproduction are released.

\section{Experimental Setup}\label{sec:setup}
Two single-thread hardware points: \textbf{x86} (Intel Core i9-9980HK, AVX2, DDR4;
$\beta_{\text{scan}}\approx 14$--$16$\,GB/s) and \textbf{M4} (Apple~M4, NEON, unified memory;
$\beta_{\text{scan}}\approx 70$, $\beta_{\text{triad}}\approx 93$\,GB/s). $N=67{,}108{,}864$ values,
median of 7 repeats, \texttt{-O3 -march=native}. M4 is treated as a \emph{held-out generalization}
point: Apple silicon does not expose reliable turbo control or hard core-pinning, so we use it to test
whether the law's \emph{form} transfers, not as a precision gate.

\section{Results: the law}\label{sec:results}
\begin{figure}[h]\centering
\includegraphics[width=0.49\textwidth]{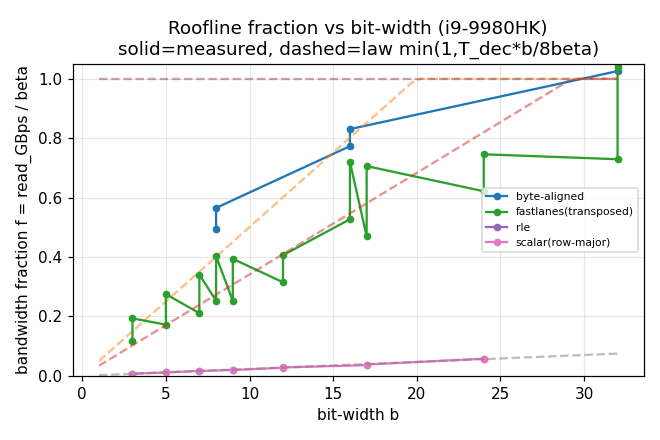}\hfill
\includegraphics[width=0.49\textwidth]{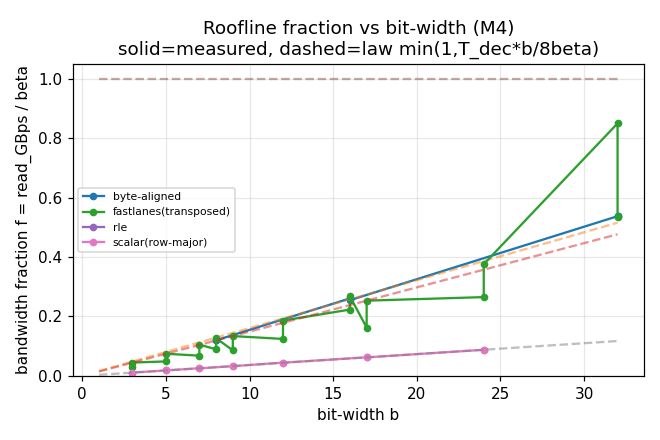}
\caption{Bandwidth fraction $f$ vs bit-width $b$ (solid = measured, dashed = law
$\min(1,\Tdec b/8\beta)$). Left: x86/AVX2. Right: Apple~M4/NEON (held-out). Scalar row-major is
compute-bound at all real widths; the transposed and byte-aligned strategies cross into
bandwidth-bound at $\bstar=8\beta/\Tdec$.}
\label{fig:roofline}
\end{figure}

Figure~\ref{fig:roofline} plots measured $f$ versus $b$ with the law overlaid. For every strategy the
measured value-throughput $\Tdec$ is flat across bit-width, and $f$ rises linearly until it plateaus
at the ridge $\bstar=8\beta/\Tdec$. Fitting one $\Tdec$ per (machine, strategy) gives the parameters
in Table~\ref{tab:fit}.

\begin{table}[h]\centering
\begin{tabular}{llrrr}
\toprule
machine & strategy & $\Tdec$ (Gval/s) & $\bstar$ (bits) & median $|f-\hat f|$ \\
\midrule
x86 & scalar (row-major)      & 0.27 & 426 & 0.0005 \\
x86 & FastLanes (transposed)  & 3.91 & 29  & 0.089 \\
x86 & byte-aligned            & 5.74 & 20  & 0.040 \\
M4  & scalar (row-major)      & 2.06 & 273 & 0.0002 \\
M4  & FastLanes (transposed)  & 8.36 & 67  & 0.022 \\
M4  & byte-aligned            & 9.04 & 62  & 0.008 \\
\bottomrule
\end{tabular}
\caption{One $\Tdec$ per strategy fits the bandwidth fraction. Overall median $|f-\hat f|$:
\textbf{0.027} on x86 (training) and \textbf{0.003} on M4 (held-out). The ridge $\bstar$ moves with
the machine: M4's $\sim\!4\times$ bandwidth pushes it from $b\!\approx\!29$ to $b\!\approx\!67$.}
\label{tab:fit}
\end{table}

\paragraph{Reproducing SOTA as a limit.} FastLanes reports $\sim\!140$ Gval/s scalar-auto-vectorized
decode and the headline that 3-bit unpacking already saturates RAM bandwidth
($\sim\!52$\,GB/s)~\citep{afroozeh2023fastlanes}. Plugging $\Tdec=140$ into Eq.~\eqref{eq:law} gives
$\bstar\approx 1$ bit and $f\approx 1$ even at $b=3$---exactly their ``decode is free'' result. The
same law that describes our portable decoder ($\Tdec\!\approx\!4$--$9$, ridge at $b\!\approx\!20$--$67$)
predicts the SOTA regime by changing one number.

\section{Results: the two crossovers}\label{sec:cross}
\begin{figure}[h]\centering
\includegraphics[width=0.49\textwidth]{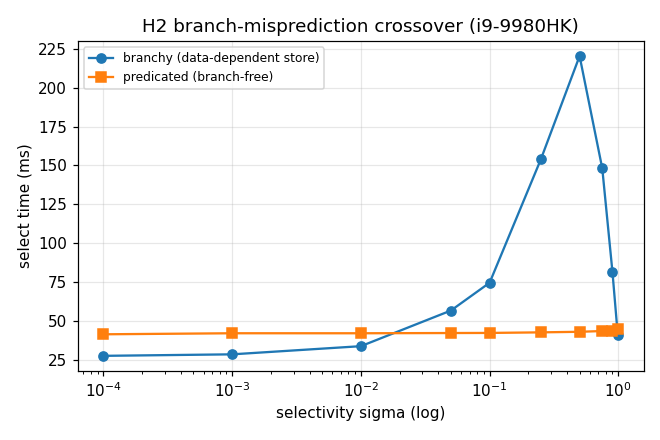}\hfill
\includegraphics[width=0.49\textwidth]{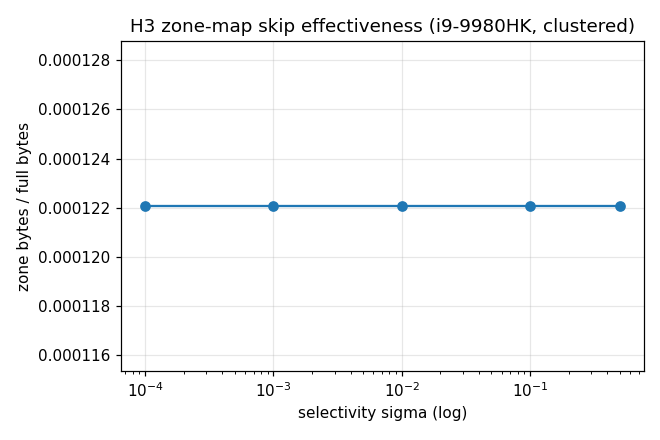}
\caption{Left: branchy select time is the $\sigma(1-\sigma)$ misprediction parabola; predicated is
flat (x86). Right: on sorted/clustered data zone maps read a tiny fraction of bytes; on random data
they do not (x86).}
\label{fig:cross}
\end{figure}

\paragraph{Branch (Fig.~\ref{fig:cross} left).} On a decode-free materializing select, branchy
evaluation time traces the textbook $\sigma(1-\sigma)$ misprediction parabola (peak at $\sigma=0.5$)
while branch-free (predicated) is flat. Branch-free wins in a mid-selectivity band; fitting
$A+B\,\sigma(1-\sigma)$ against the constant predicated cost predicts the band edges within $\pm1$
selectivity decade on both machines (x86: model $(0.026,0.974)$ vs measured $[0.05,0.90]$; M4:
$(0.008,0.992)$ vs $[0.001,0.99]$). Caveat: at \texttt{-O3} the compiler auto-branchlesses a
\emph{count} aggregate, collapsing the distinction---the crossover is real only for data-dependent
(materializing / position-list) scans.

\paragraph{Clustering (Fig.~\ref{fig:cross} right).} Zone-map skipping is \emph{clustering}-gated, not
selectivity-gated. On a sorted column the scan reads only the block(s) overlapping the predicate value
($\sim\!1/8192$ of bytes here), so skipping wins at every selectivity; on random data zone maps skip
nothing (bytes read $\approx$ full) except at $\sigma$ below one match per block. This sharpens the
common ``zone maps help only at low selectivity''~\citep{zeng2023empirical}: the lever is whether the
predicate value is confined to few blocks.

\FloatBarrier
\section{Related Work}
Vectorized, bandwidth-aware scans originate with MonetDB/X100~\citep{boncz2005monetdb}; Data
Blocks~\citep{lang2016datablocks}, BtrBlocks~\citep{kuschewski2023btrblocks}, and
FastLanes~\citep{afroozeh2023fastlanes} push decode toward the roofline with SIMD-friendly encodings.
\citet{polychroniou2015lightweight,polychroniou2015rethinking} study branch-free SIMD scans;
\citet{zeng2023empirical} empirically vary selectivity, encoding, and zone-map effectiveness;
\citet{manegold2002generic} model memory-access cost (latency, pre-roofline);
\citet{kersten2018everything} compare compiled vs vectorized execution qualitatively. Our delta is the
closed-form bandwidth-fraction law $f=\min(1,\Tdec b/8\beta)$, the ridge $\bstar=8\beta/\Tdec$, its
cross-hardware validation, and the reproduction of SOTA as the large-$\Tdec$ limit. We claim none of
the \emph{phenomena}---selectivity dependence, the branch crossover, zone-map skipping---only the
predictive model that unifies them.

\section{Limitations and Honest Negatives}\label{sec:limitations}
\textbf{(1) Absolute throughput.} Our portable-C decoders reach $\sim\!2$ values/cycle, $\sim\!5\%$ of
FastLanes' hand-tuned $\sim\!40$; we did not reimplement the full 1024-bit interleaving and
instruction scheduling. The law is parameterized by the measured $\Tdec$, so this does not affect the
fit, but the absolute ridge $\bstar$ we report is specific to our decoders (and the law predicts how
$\bstar$ moves for a faster one). \textbf{(2) Single-thread.} We model per-core bandwidth; the
many-core shared-controller (aggregate) wall is out of scope. \textbf{(3) Counting vs materializing.}
The branch crossover requires a data-dependent kernel; for count aggregates the compiler removes the
branch. \textbf{(4) Integer columns.} Strings/nested types (where SOTA decode is hardest) are future
work; we expect the law to hold with their $\Tdec$. \textbf{(5) Measurement.} Apple-silicon lacks
turbo/pinning control, so M4 is a generalization point, not a precision gate; we report medians.

\section{Conclusion}
A columnar scan's bandwidth fraction is captured by one number per decode strategy: its
bit-width-independent value throughput $\Tdec$, via $f=\min(1,\Tdec b/8\beta)$ with ridge
$\bstar=8\beta/\Tdec$. The law fits across two ISAs, predicts where decode stops being free, and
reproduces hand-tuned SOTA as a limit. With the branch and clustering crossovers and an open harness,
it gives practitioners a direct answer to ``will this scan be bandwidth-bound, and at what bit-width?''

\paragraph{Artifact.} Code, data, and one-command reproduction:
\url{https://github.com/samyama-ai/bandwidth-bound-scan}.

\small
\bibliographystyle{plainnat}
\bibliography{paper14_bandwidth_scan}
\end{document}